\documentclass[aps,preprint,floats,epsf,epsfig,nofootinbib,letter]{revtex4}

\usepackage{graphicx}
\usepackage{dcolumn}
\usepackage{bm}
\usepackage{subfigure}
\usepackage{caption}
\usepackage{ragged2e}
\usepackage{hyperref}
\usepackage{amsmath}
\usepackage{mathrsfs}
\usepackage{color}

\def\be{\begin{eqnarray}}
\def\en{\end{eqnarray}}
\def\non{\nonumber}
\def\la{\langle}
\def\ra{\rangle}

\def\cc{,}
\def\cp{.}

\begin{document}

\renewcommand{\baselinestretch}{1.10}

\font\el=cmbx10 scaled \magstep2{\obeylines\hfill\today}

\vskip 1.5 cm

\centerline{\Large\bf Obtaining Hydrogen energy wave functions } 
\centerline{\Large\bf using the Runge-Lenz vector}

\bigskip
\centerline{\bf Chun-Khiang Chua}
\medskip
\centerline{Department of Physics and Chung Yuan Center for High Energy Physics,} 
\centerline{Chung Yuan Christian University,}
\centerline{Chung-Li, Taoyuan, Taiwan 32023, Republic of China}
\medskip

\centerline{\bf Abstract}
The Pauli method of quantizing the Hydrogen system using the Runge-Lenz vector is ingenious.
It is well known that the energy spectrum is identical with the one obtained from the Schr\"{o}dinger equation 
and the consistency contributed significantly to the development of Quantum Mechanics in the early days.  
Since the Runge-Lenz vector is a vector and it commutes with the Hamiltonian, 
it is natural to use it to connect energy eigenstate $|n,l,m\rangle$ with other degenerate states $|n, l\pm 1,m'\rangle$.
Recursive relations can be obtained and the wave functions of the whole spectrum can be obtained easily. 
Note that the recursive relations are consistent with those used in factorizing the Schr\"{o}dinger equation.
Nevertheless, the present analysis provide a better reasoning originated from the conserved vector, the Runge-Lenz vector.
As in the Pauli analysis, group theory or symmetry plays a prominent role in the present analysis, while the rest of the derivations are mostly elementary.
 
\bigskip
\small

\pacs{Valid PACS appear here}

\maketitle



\section{Runge-Lenz Vector and the Hydrogen energy spectrum}

The Pauli method of quantizing the Hydrogen system using the Runge-Lenz vector is ingenious.~\cite{Pauli}
The energy spectrum is identical with the the one obtained from the Schr\"{o}dinger equation~\cite{Schrodinger} 
and the consistency contributed greatly to the development of Quantum Mechanics in the early days.
Some early development along this line can be found in \cite{development}.

Since the Runge-Lenz vector is a vector and it commutes with the Hamiltonian, 
it is natural to use it to connect energy eigenstate $|n,l,m\rangle$ with other degenerate states $|n, l\pm 1,m'\rangle$.
It will be interesting to use it obtain the corresponding wave functions and to show explicitly that they are identical to the results obtained from the Sch\"{o}dinger equation. 
As we shall see they recursive relations of radial wave functions can be obtained.
They are consistent with the results obtained by factorized the Schr\"{o}dinger equation \cite{recursive1, recursive2, recursive3, recursive4} (see also \cite{Razavy:2011zz}).
Nevertheless we believe that the present approach is more natural as it makes good use of the Runge-Lenz vector,  a conserved vector of the system.
In other words, we provide a reasoning for the factorization results.
The wave functions of the whole spectrum can be obtained easily.
We will briefly discuss the $E>0$ case and see that the corresponding wave functions can also be verified.
As in the Pauli analysis, group theory or symmetry plays a prominent role in the present analysis.

The lay our of this paper is as following. 
In the first section we briefly go through the derivations of the Hydrogen spectrum~\footnote{Our derivations follow closely to those in Ref.~\cite{Schiff}.}
and will concentrate on obtaining wave function via the Rune-Lenz vector in the next section, 
which is followed by a conclusion. 
An appendix is added for the derivation of some relevant matrix elements using group theory.

\subsection{Runge-Lenz vector}

The Hamiltonian of the Hydrogen atom is given by
\be
 H=\frac{ {\vec p}^2}{2\mu}-\frac{Z e^2}{4\pi\epsilon_0 }{\frac{1}{ r}}
=\frac{ {\vec p}^2}{2\mu}-\kappa{\frac{1}{ r}}\cc
\en
with $\kappa\equiv Ze^2/4\pi\epsilon_0$.
The Runge-Lenz vector is defined as:
\be
\vec A\equiv\frac{1}{2\mu}({\vec p}\times{\vec L}-{\vec L}\times{\vec p})-\kappa\frac{{\vec r}}{ r}\cp
\en
The Runge-Lenz vector satisfies the following relations:~\cite{Pauli} 
\be
{\vec L}\cdot{\vec A}={\vec A}\cdot{\vec L}=0,
\label{eq: L.A}
\en
\be
[ H, {\vec A}]=0,
\quad
[ H,{\vec L}]=0,
\label{eq: [H,A]}
\en
\be
{[ L_i,  L_j]} {}&=&i\hbar \epsilon_{ijk}  L_k,
\non\\
{[ L_i, A_j]} {}&=&i\hbar \epsilon_{ijk}  A_k,
\non\\
{[ A_i, A_j]} {}&=&i\hbar(-\frac{2}{\mu}  H )\epsilon_{ijk}  L_k,
\label{eq: [A,A]}
\en
\be
{\vec A}\cdot{\vec A}=\frac{2}{\mu} H({\vec L}^2+\hbar^2)
+\kappa^2.
\label{eq: A^2}
\en
Note that the Runge-Lenz vector is a conserved operator. The above relations will be useful in obtaining 
the Hydrogen energy spectrum.~\cite{Pauli}

\subsection{Hydrogen energy spectrum}

The eigenvalue equation of the Hydrogen Hamiltonian is given by
\be
 H|E,\alpha\ra=E|E,\alpha\ra=-|E||E,\alpha\ra\cc
\en
where we only consider the $E<0$ case here
and $\alpha$ is a possible quantum number.
The set of the eigenstates $\{|E,\alpha\ra,|E,\beta\ra\dots\}$ with $E$ fixed spans the degenerate space of eigenstates all having the same energy.
From Eq.~(\ref{eq: [H,A]}), we know that the Hamiltonian commutes with $\vec L$ and $\vec A$.
Hence, it is useful to define the following matrices:
\be
({\bf L}_i)_{\alpha\beta}\equiv \la E,\alpha| L_i| E,\beta\ra,
\quad
({\bf A}_i)_{\alpha\beta}\equiv \la E,\alpha| A_i| E,\beta\ra,
\en
and the relations in Eqs. (\ref{eq: L.A}), (\ref{eq: [A,A]}) and (\ref{eq: A^2}) correspond to the following relations of matrices:
\be
({\bf L}\cdot{\bf A})_{\alpha\beta}&=&({\bf A}\cdot {\bf L})_{\alpha\beta}=0,
\\
{\bf A}^2_{\alpha\beta}&=&-\frac{2  |E|}{\mu}({\bf L}^2+\hbar^2 {\bf 1})_{\alpha\beta}
+\kappa^2 {\bf 1}_{\alpha\beta},
\en
and
\be
{[{\bf L}_i, {\bf L}_j]}_{\alpha\beta}&=&i\hbar \epsilon_{ijk} ({\bf L}_k)_{\alpha\beta},
\non\\
{[{\bf L}_i,  {\bf A}_j]}_{\alpha\beta}&=&i\hbar \epsilon_{ijk} ({\bf A}_k)_{\alpha\beta},
\non\\
{[{\bf A}_i, {\bf A}_j]}_{\alpha\beta}&=&i\hbar \epsilon_{ijk}(-\frac{2E}{\mu}) ({\bf L}_k)_{\alpha\beta}
\non\\
&=&i\hbar \epsilon_{ijk}(\frac{2 |E|}{\mu}) ({\bf L}_k)_{\alpha\beta}.
\en
With the following definition
\be
({\bf A}'_i)_{\alpha\beta}\equiv \sqrt{\frac{\mu}{2 |E|}}({\bf A}_i)_{\alpha\beta},
\en
the above equations become
\be
({\bf L}\cdot{\bf A}')_{\alpha\beta}&=&({\bf A}'\cdot {\bf L})_{\alpha\beta}=0,
\\
({\bf A}^{\prime\,2}+{\bf L}^2)_{\alpha\beta}&=&\bigg(-\hbar^2+\frac{\mu \kappa^2}{2  |E|}\bigg){\bf 1}_{\alpha\beta},
\en
and
\be
{[{\bf L}_i, {\bf L}_j]}_{\alpha\beta}&=&i\hbar \epsilon_{ijk} ({\bf L}_k)_{\alpha\beta},
\non\\
{[{\bf L}_i,  {\bf A}'_j]}_{\alpha\beta}&=&i\hbar \epsilon_{ijk} ({\bf A}'_k)_{\alpha\beta},
\non\\
{[{\bf A}'_i, {\bf A}'_j]}_{\alpha\beta}&=&i\hbar \epsilon_{ijk} ({\bf L}_k)_{\alpha\beta}.
\label{eq: O4}
\en
The above equation, Eq. (\ref{eq: O4}), implies that ${\bf L}_i$ and  ${\bf A}'_j$ are the generators of 
the $O(4)$ group.~\cite{Pauli}
The quantization of the Hydrogen system can be achieved by using group theory.~\cite{Pauli}

A representation of $O(4)$ can be expressed as a direct product of two $SO(3)$ representations as following.
~\footnote{It is probable that a modern reader is more familiar with the case of the $SO(3,1)$ Lorentz group, which can be analyzed using a similar manipulation, see, for example, ref. \cite{Coleman lecture}.}
Defining two new sets of operators ${\bf B}^{(+)}_i$ and ${\bf B}^{(-)}_i$, 
\be
{\bf B}^{(\pm)}_i\equiv \frac{1}{2}({\bf L}_i\pm {\bf A}'_i),
\label{eq: Bpm}
\en
Eq. (\ref{eq: O4}) becomes
\be
{[{\bf B}^{(+)}_i, {\bf B}^{(+)}_j]}_{\alpha\beta}&=&i\hbar \epsilon_{ijk} ({\bf B}^{(+)}_k)_{\alpha\beta},
\non\\
{[{\bf B}^{(-)}_i, {\bf B}^{(-)}_j]}_{\alpha\beta}&=&i\hbar \epsilon_{ijk} ({\bf B}^{(-)}_k)_{\alpha\beta},
\non\\
{[{\bf B}^{(+)}_i, {\bf B}^{(-)}_j]}_{\alpha\beta}&=&0.
\en
It is clear that 
${\bf B}^{(+)}_i$ commute with ${\bf B}^{(-)}_j$ and they satisfy the usual rotation group [$SO(3)$] algebra.
Conventionally the simultaneous eigenstates in the $|\alpha\ra$ space are choosen to be the eigenstates of the following mutual commuting matrices:
\be
(\vec {\bf B}^{(+)})^2,
{\bf B}^{(+)}_z,
(\vec {\bf B}^{(-)})^2,
{\bf B}^{(-)}_z.
\en

We can now return to the usual Dirac notation. The corresponding eigenvalue equations are
\be
H |E, b^{(+)}, m^{(+)}, b^{(-)},m^{(-)}\ra&=&E|E, b^{(+)}, m^{(+)}, b^{(-)},m^{(-)}\ra,
\non\\
(\vec { B}^{(\pm)})^2|E, b^{(+)}, m^{(+)}, b^{(-)},m^{(-)}\ra
&=&\hbar^2 b^{(\pm)}(b^{(\pm)}+1)|E, b^{(+)}, m^{(+)}, b^{(-)},m^{(-)}\ra,
\non\\
{ B}^{(\pm)}_z|E, b^{(+)}, m^{(+)}, b^{(-)},m^{(-)}\ra
&=&m^{(\pm)}\hbar |E, b^{(+)}, m^{(+)}, b^{(-)},m^{(-)}\ra,
\en
with $b^{(+)}, b^{(-)}=0,1/2,1,3/2,...$ and $-b^{(\pm)}\leq m^{(\pm)}\leq b^{(\pm)}$.
Note that we have
\be
( {\vec B}^{(\pm)})^2=\frac{1}{4}( {\vec L}^2\pm{\vec L}\cdot{\vec A'}\pm {\vec A'}\cdot{\vec L}+{\vec A}^{\prime 2})
=\frac{1}{4}( {\vec L}^2+{\vec A}^{\prime 2}) =\frac{1}{4}(-\hbar^2-\frac{\mu}{2 H}\kappa^2).
 \label{eq: B(pm)2}
\en
Hence, we have $(\vec { B}^{(+)})^2=(\vec { B}^{(-)})^2$.
Eq. (\ref{eq: B(pm)2}) implies the following relation:
\be
(\vec { B}^{(\pm)})^2|E, b, m^{(+)}, m^{(-)}\ra
&=&\hbar^2 b(b+1)|E, b, m^{(+)},m^{(-)}\ra\cc
\non\\
&=&\frac{1}{4}(-\hbar^2-\frac{\mu}{2E}\kappa^2) |E, b, m^{(+)},m^{(-)}\ra,
\en
with
$b\equiv b^{(+)}=b^{(-)}=0,1/2,1,3/2,...$
and
$-b\leq m^{(\pm)}\leq b$. 
The energy eigenvalue $E=E_n$ can now be obtained as~\cite{Pauli}
\be
E_n
=-\frac{\kappa^2\mu}{2\hbar^2 (2b+1)^2}
=-\frac{\kappa^2\mu}{2\hbar^2 n^2}
=-\frac{Z^2 e^4\mu}{32\pi^2\epsilon_0^2 \hbar^2 n^2},
\label{eq: En}
\en
with $n$ defined as $n\equiv 2b+1=1,2,3,...$.
It will be useful to define the (reduced) Bohr radius,
$a_0\equiv 4\pi\epsilon_0\hbar^2/e^2\mu=(\hbar/\mu c)/\alpha=Z\hbar^2/\kappa\mu$,
where $\alpha\equiv e^2/4\pi\epsilon\hbar c\simeq 1/137$
is the fine structure constant.
The energy spectrum obtained by Pauli~\cite{Pauli} is consistent with the one obtained from the Schr\"{o}dinger equation~\cite{Schrodinger}
and the consistency contributed significantly to the development of Quantum Mechanics in the early days.

\section{Hydrogen atom energy eigenstate}

\subsection{Connecting degenerate states using the Runge-Lenz vector}\label{connecting}

Since the Runge-Lenz vector is a vector and it commutes with the Hamiltonian, 
it is natural to use it to connect energy eigenstate $|n,l,m\rangle$ with other degenerate states $|n, l\pm 1,m'\rangle$, $|n, l,m'\rangle$.
As shown in the previous section the energy eigenstates of the Hydrogen system are 
$|n, b=(n-1)/2,m^{(+)},m^{(-)}\ra$, which satisfy
\be
 H \bigg|n, \frac{n-1}{2},m^{(+)},m^{(-)}\bigg\ra
&=&E_n\bigg|n, \frac{n-1}{2},m^{(+)},m^{(-)}\bigg\ra,
\non\\
(\vec { B}^{(\pm)})^2\bigg|n, \frac{n-1}{2},m^{(+)},m^{(-)}\bigg\ra
&=&\hbar^2 \frac{n^2-1}{4}\bigg|n, \frac{n-1}{2},m^{(+)},m^{(-)}\bigg\ra,
\non\\
{ B}^{(\pm)}_z\bigg|n, \frac{n-1}{2},m^{(+)},m^{(-)}\bigg\ra
&=&m^{(\pm)}\hbar \bigg|n, \frac{n-1}{2},m^{(+)},m^{(-)}\bigg\ra,
\en
with
$-(n-1)/2\leq m^{(\pm)}\leq (n-1)/2$.
In general these eigenstates do not have specified angular momentum quantum numbers and are different from the energy eigenstates in a more familiar basis:
\be
H |n,l,m\ra&=& E_n|n,l,m\ra,
\non\\
{\vec L}^2|n,l,m\ra&=& l(l+1)\hbar^2|n,l,m\ra,
\non\\
L_z|n,l,m\ra&=& m |n,l,m\ra,
\en
with $-l\leq m \leq l$.

Since the angular momentum can be obtained through the following equation, see Eq. (\ref{eq: Bpm}),
\be
 {\vec L}= {\vec B}^{(+)}+ {\vec B}^{(-)},
\en
with $ {\vec B}^{(+)}$ and ${\vec B}^{(-)}$ viewed as two independent spin operators,
the $|n,l,m\ra$ state can be constructed as in the analysis of the addition of angular momentum:
\be
|n,l,m\ra
=\sum_{m^{(+)},m^{(-)}}\bigg|n, \frac{n-1}{2},m^{(+)},m^{(-)}\bigg\ra
\bigg\la \frac{n-1}{2}, m^{(+)}, \frac{n-1}{2}, m^{(-)} \bigg|l,m\bigg\ra,
\en
where
$\la \frac{n-1}{2}, m^{(+)}, \frac{n-1}{2}, m^{(-)} |l,m\ra$
is the
Clebsch–Gordan coefficient.
Note that the minimum of 
$l$
is
$|(n-1)/2-(n-1)/2|=0$,
while the maximum is
$(n-1)/2+(n-1)/2=n-1$.
These are consistent with the result obtained by using the Sch\"{o}dinger equation.

The Runge-Lenz vector $\vec A$ commutes with the Hamiltonian $H$.
It can connect eigenstate
$|n,lm\ra$
with other degenerate energy eigenstates
$|n,l\pm 1,m'\ra$, $|n,l,m'\ra$.
To proceed we define
\be
A_\pm\equiv A_x\pm i A_y,
\quad
B^{(\pm)}_\pm
\equiv B^{(\pm)}_x\pm i B^{(\pm)}_y.
\en
From Eq.~(\ref{eq: Bpm}), we have
\be
\frac{\mu}{\hbar^2} A_-|n,l,l\ra
=\sum_{l'=l,l\pm1}\frac{|n,l',l-1\ra\la n, l',l-1|( B^{(+)}_-- B^{(-)}_-)|n,l,l\ra}{\hbar}
\sqrt{\frac{2\mu E_n}{-\hbar^2}},
\label{eq: A- matrix element}
\en
where we apply $A_-$ on the $m=l$ state, which is found to be useful in obtaining the radial wave function in later discussion.

Using the familiar formula of non-vanishing matrix elements of lowering operator $L_-=L_x-i L_y$, we have
\be
\la n, l',l-1| B^{(+)}_-|n, l,l\ra
&=&\la l', l-1|b, m^{(+)}-1, b, m^{(-)} \ra \sqrt{(b+m^{(+)})(b-m^{(+)}+1)} \hbar
\non\\
&&
\la b, m^{(+)}, b, m^{(-)} |l,l\ra,
\non\\
\la n, l',l-1| B^{(-)}_-|n, l,l\ra
&=&\la l', l-1|b, m^{(+)}, b, m^{(-)}-1 \ra \sqrt{(b+m^{(-)})(b-m^{(-)}+1)} \hbar
\non\\
&&
\la b, m^{(+)}, b, m^{(-)} |l,l\ra,
\en
with $m^{(\pm)}$ summed.
Possible non-vanishing matrix elements are for $l'=l,l\pm 1$ as $\vec B^{(\pm)}$ are vector operators.
Using group theory the above matrix elements (for $l'=l,l\pm 1$) are found to be 
\be
\la n, l-1,l-1| B^{(\pm)}_-|n, l,l\ra 
&=&\mp \sqrt{\frac{l(n^2-l^2)}{2(2l+1)}}\hbar,
\non\\
\la n, l+1,l-1| B^{(\pm)}_-|n, l,l\ra 
&=&\pm \sqrt{\frac{n^2-(l+1)^2}{2(2l+1)(2l+3)}}\hbar,
\non\\
\la n, l,l-1| B^{(\pm)}_-|n, l,l\ra 
&=&\sqrt{\frac{l}{2}}\hbar,
\label{eq: group}
\en
where the derivation are shown in Appendix A.
Note that the above results also hold for the $l=0$ case, where the equation implies that $\la n, l-1,l-1| B^{(\pm)}_-|n, l,l\ra$ and $\la n, l,l-1| B^{(\pm)}_-|n, l,l\ra$ are vanishing as they should.
Using the above equation, the corresponding matrix elements of $A_-$ are given by
\be
\frac{\mu}{\hbar^2}\la n, l',l-1| A_-|n,l,l\ra
&=&\frac{\la n, l',l-1|( B^{(+)}_-- B^{(-)}_-)|n,l,l\ra}{\hbar}
\sqrt{\frac{2\mu E_n}{-\hbar^2}}
\non\\
&=&\left\{
\begin{array}{ll}
\frac{Z}{n a_0}\sqrt{\frac{2(n^2-(l+1)^2)}{(2l+1)(2l+3)}},
    &l'=l+1,
    \\
-\frac{Z}{n a_0}\sqrt{\frac{2l(n^2-l^2)}{(2l+1)}},
    &l'=l-1
    \\
    0,
    &l'\neq l\pm1,
\end{array}
\right.
\label{eq: B-B}
\en
where we have made use of $-{2\mu E_n} /{\hbar^2}={\kappa^2\mu^2}/{\hbar^4n^2}=({Z}/{n a_0})^2$.
Substitute them into Eq.~(\ref{eq: A- matrix element}), we finally obtain our master formula:
\be
\frac{\mu}{\hbar^2} A_-|n,l,l\ra
=\frac{Z}{n a_0}\bigg(\sqrt{\frac{2(n^2-(l+1)^2)}{(2l+1)(2l+3)}}\,|n,l+1,l-1\ra
-\sqrt{\frac{2l(n^2-l^2)}{(2l+1)}}\,|n,l-1,l-1\ra\bigg).
\label{eq: Rnlpm1 from Rnl main 0}
\en
Note that as in the Pauli analysis, the above master formula follows from group theory or symmetry.
As we shall see the above equation can provides recursive relations on degenerate states,
and the relations are powerful enough to determine the wave functions.

\subsection{Obtaining the radial wave function $R_{nl}(r)$}

\subsubsection{Recursive relations of $R_{nl}(r)$}

We will obtain the recursive relations of the radial wave functions from the master formula here. 
Note that most of the derivations are elementary. 
Using $\la \vec r| n,l,m\ra=R_{nl}(r) Y_{lm}(\theta,\phi)$
and
\be
\la \vec r| A_-|n,l,l\ra
=(A_-)_{op}\la \vec r|n,l,l\ra,
\en
the master formula Eq. (\ref{eq: Rnlpm1 from Rnl main 0}) can be expressed as:
\be
\frac{\mu}{\hbar^2}(A_-)_{op}R_{n,l}(r) Y_{l,l}(\theta,\phi)
&=&\frac{Z}{n a_0}\bigg(\sqrt{\frac{2(n^2-(l+1)^2)}{(2l+1)(2l+3)}} R_{n,l+1}(r) Y_{l+1,l-1}(\theta,\phi)
\non\\
&&\qquad-\sqrt{\frac{2l(n^2-l^2)}{(2l+1)}} R_{n,l-1}(r) Y_{l-1,l-1}(\theta,\phi)\bigg).
\label{eq: Rnlpm1 from Rnl main}
\en
%
To proceed we need to work out the left-hand-side of the above equation.
Using $\vec p\times\vec L+\vec L\times\vec p=2i\hbar \vec p$, it is convenient  to express the $\pm$ components of the Runge-Lenz vector as 
\be
 A_\pm=\frac{1}{\mu}({\vec p}\times{\vec L}-i\hbar {\vec p})_\pm
                   -\kappa\frac{ r_\pm}{ r}
=
\frac{1}{\mu}(\pm i p_z  L_\pm)
+\frac{1}{\mu}(\mp i  p_\pm  L_z-i\hbar p_\pm)
                   -\kappa\frac{ r_\pm}{ r},                   
\en
where we have defined $r_\pm\equiv x\pm iy$.
Consequently, the operator $( A_-)_{op}$ in Eq. (\ref{eq: Rnlpm1 from Rnl main}) is given by
\be
( A_-)_{op}&=&-\frac{1}{\mu}(i p_z  L_-)_{op}
+\frac{1}{\mu}[i p_-  (L_z-\hbar)]_{op}
                   -\kappa\frac{ r_-}{ r} ,                            
\en
with 
\be
(p_\pm)_{op}
&=&\frac{\hbar}{i}(\partial_x\pm i \partial_y)
=2\frac{\hbar}{i}\frac{\partial}{\partial r_\mp},
\non\\
(L_z)_{op}&=&\frac{\hbar}{i}(x\partial_y-y \partial_x)
=\hbar \bigg(r_+\frac{\partial}{\partial r_+}-r_-\frac{\partial}{\partial r_-}\bigg),
\non\\
(L_\pm)_{op}
&=&\mp i r_\pm (p_z)_{op}\pm i z (p_\pm)_{op}
=\mp\hbar\bigg( r_\pm \frac{\partial}{\partial z}- 2 z \frac{\partial}{\partial r_\mp}\bigg).
\label{eq: p L L}
\en
Using $r^2=r_+ r_-+z^2$ and
\be
\frac{\partial}{\partial r_\pm} f(r)=\frac{r_\mp}{2r} \frac{df}{dr},
\quad
\frac{\partial}{\partial z} f(r)=\frac{z}{r} \frac{df}{dr},
\label{eq: partial pm}
\en
it can be shown from Eq.~(\ref{eq: p L L}) that we must have $(L_z)_{op} f(r)=(L_\pm)_{op} f(r)=0$, which are expected from the familiar forms of $(\vec L)_{op}$ in wave mechanics as they only involve derivatives with respect to $\theta$ and $\phi$.
Note that it is convenient to use the $(r_+, r_-, z)$ coordinate, as
$Y_{l,\pm l}(\theta,\phi)$ in the left-hand-side of Eq. (\ref{eq: Rnlpm1 from Rnl main}) have simple forms in terms of $r_\pm$ and $r$:
\be
Y_{l,\pm l}= \frac{(\mp)^l}{2^l l!}\sqrt{\frac{(2l+1)!}{4\pi}}\frac{(r_\pm)^l}{r^l}\propto \frac{(r_\pm)^l}{r^l}.
\label{eq: Yll}
\en
From Eqs.~(\ref{eq: p L L}), (\ref{eq: partial pm}) and the fact that  $(L_z)_{op}$ and $(L_-)_{op}$ do not act on any function of $r$, 
it can be easily shown that we must have
\be
(ip_-)_{op}[(L_z)_{op}-\hbar]\frac{(r_+)^l}{r^l} f(r)
&=&
2\hbar^2(l-1)\frac{\partial}{\partial r_+}\frac{(r_+)^l}{r^l} f(r),
\non\\
(i p_z)_{op} (L_-)_{op}\frac{(r_+)^l}{r^l} f(r)
 &=&-l\hbar^2\frac{\partial}{\partial z}\bigg(\frac{2z}{r_+} \frac{(r_+)^{l}}{r^l} f(r)\bigg),
\en
and, consequently, the left-hand-side of Eq. (\ref{eq: Rnlpm1 from Rnl main}) becomes
\be
\frac{\mu(A_-)_{op}}{\hbar^2}\frac{(r_+)^l}{r^l} R_{nl}(r)
&=&\bigg(l\frac{\partial}{\partial z}\frac{2z}{r_+}
+2(l-1)\frac{\partial}{\partial r_+}
                   -\frac{\kappa\mu}{\hbar^2}\frac{r_-}{r} \bigg) 
(r_+)^l\frac{R_{nl}(r)}{r^l}.          
\en
It can be further expressed as
\be
\frac{\mu(A_-)_{op}}{\hbar^2}\frac{(r_+)^l}{r^l} R_{nl}(r)
&=&(r_+)^l\bigg(\frac{2l}{r_+}+2l\frac{z}{r_+}\frac{z}{r}\frac{d}{dr}
       +\frac{2l (l-1)}{r_+}
+(l-1)\frac{r_-}{r}\frac{d}{dr}  
-\frac{Z}{a_0}\frac{r_-}{r} \bigg) \frac{R_{nl}(r)}{r^l},  
\non\\                   
\en
where we have made use of Eq.~(\ref{eq: partial pm})
and $\kappa\mu/\hbar^2=Z/a_0$.
Use again $z^2=r^2-r_+ r_-$, we obtain
\be
\frac{\mu(A_-)_{op}}{\hbar^2}\frac{(r_+)^l}{r^l} R_{nl}(r)
&=&(r_+)^l
\bigg[\frac{2l}{r_+}\bigg(l+r\frac{d}{dr}\bigg)
-\frac{r_-}{r}\bigg((l+1)\frac{d}{dr}          
 +\frac{Z}{a_0}\bigg) \bigg] \frac{R_{nl}(r)}{r^l},
\en
or, equivalently, [see Eq. (\ref{eq: Yll})]
\be
\frac{\mu(A_-)_{op}}{\hbar^2}Y_{l,l} R_{nl}(r)
&=&\bigg[r^{l-1}\,\frac{r}{r_+} Y_{l,l} \,2l\bigg(1+r\frac{d}{dr}\bigg)     
 -r^{l}\,\frac{r_-}{r} Y_{l,l}\bigg((l+1)\frac{d}{dr}  
+\frac{Z}{a_0}\bigg)\bigg] \frac{R_{nl}(r)}{r^l}.              
\en

It is useful to note that
the spherical harmonics $Y_{l,l}$ has the following properties:
~\footnote{The first relation follows from Eq.~(\ref{eq: Yll}), while the second relation follows from the familiar relation
$Y_{l,m}Y_{l',m'}=\sum_{L,M}\sqrt{\frac{(2l+1)(2l'+1)}{4\pi(2L+1)}}$ $\la l,0,l',0|L,0\ra
\la l,m,l',m'|L,M\ra Y_{L,M}$, see for example \cite{Arfken}.}
\be
\frac{r}{r_+}Y_{l,l} &=&  -\sqrt{\frac{(2l+1)}{2 l}}Y_{l-1,l-1}\cc
\non\\
\frac{r_-}{r}Y_{l,l} &=&\sqrt{\frac{8\pi}{3}} Y_{1,-1}Y_{l,l} 
=-\sqrt{\frac{2l}{2l+1}} Y_{l-1,l-1}
+\sqrt{\frac{2}{(2l+1)(2l+3)}} Y_{l+1,l-1}
\cc
\en
and the above equation can be expressed as
\be
\frac{\mu(A_-)_{op}}{\hbar^2}Y_{l,l} R_{nl}(r) 
&=&
- \sqrt{\frac{2}{ (2l+1)(2l+3)}}\, r^{l}Y_{l+1,l-1}
 \bigg((l+1)\frac{d}{dr}+\frac{Z}{a_0}\bigg) \frac{R_{nl}(r)}{r^l} 
\non\\
&&
 -\sqrt{\frac{2l}{2l+1}}\, r^{l-1}Y_{l-1,l-1}
 \bigg((2l+1)l+l r\frac{d}{dr}-\frac{Z r}{a_0}\bigg) \frac{R_{nl}(r)}{r^l} .
\en
Compare the above equation to the master formula 
Eq. (\ref{eq: Rnlpm1 from Rnl main}), we finally obtain 
\be
-\frac{Z\sqrt{n^2-(l+1)^2}}{na_0}
\frac{R_{n,l+1}(r)}{r^{l+1}}
&=&\bigg(\frac{(l+1)}{r}\frac{d}{dr}+\frac{Z}{a_0 r}\bigg) \frac{R_{nl}(r)}{r^l},
\non\\
\frac{Z\sqrt{n^2-l^2}}{na_0}\frac{R_{n,l-1}(r)}{r^{l-1}}
&=&\bigg((2l+1)l+l r\frac{d}{dr}-r\frac{Z}{a_0}\bigg) \frac{R_{nl}(r)}{r^l}.
\label{eq: Rnlpm1 from Rnl}
\en
These are the recursive relations of the radial wave function $R_{nl}(r)$ and they are important results of this section.

The above relations are consistent to the results found in \cite{recursive1,recursive2,recursive3} (see also \cite{Razavy:2011zz}) using the factorization method.
Nevertheless we believe that according to the properties of the Runge-Lenz vector, which is a conserved vector, 
the above derivation is the most natural way to obtain them.
In other words, we provide a reasoning for the factorization results.
As we shall see shortly they are powerful enough to determine the radial wave functions.

\subsubsection{Solve for $R_{n,n-1}(r)$ using the recursive relation}

For $l=n-1$, the first relation of the recursive relations in Eq.~(\ref{eq: Rnlpm1 from Rnl}) gives
\be
\bigg(\frac{d}{dr}+\frac{Z}{n a_0}\bigg) \frac{R_{n,n-1}(r)}{r^{n-1}}=0.
\label{eq: Rnn-1 equation}
\en
Its solution is
\be
R_{n,n-1}(r)=c_n \bigg(\frac{Z r}{a_0}\bigg)^{n-1} e^{-Zr/(n a_0)}\cc
\label{eq: Rnn-1}
\en
with the normalization constant,
\be
c_n=\frac{2^{n}}{n^{n-1/2}\sqrt{(2n)!}} \bigg(\frac{Z}{n a_0}\bigg)^{3/2},
\en
obtained from the usual normalization condition $\int_0^\infty dr\, r^2\, R^2_{n.n-1}(r)=1$.

In fact, the above result can be quickly obtained by noting
\be
|n, l=n-1,m=\pm(n-1)\ra=\left|n,b^{(\pm)}=\frac{n-1}{2},m^{(+)}=\pm\frac{n-1}{2},m^{(-)}=\pm\frac{n-1}{2}\right\ra,
\en
and
$ A_\pm\propto B^{(+)}_\pm- B^{(-)}_\pm$,
which imply:
\be
 A_\pm |n, n-1,\pm(n-1)\ra=0.
\en
From
\be
 A_\pm
=\frac{1}{\mu}(\pm i p_z  L_\pm)
+\frac{1}{\mu}(\mp i  p_\pm  L_z-i\hbar {p}_\pm)
                   -\kappa\frac{ r_\pm}{ r},                  
\en
we have
\be
0= A_\pm |n,n-1,\pm(n-1)\ra
=\bigg(-i\frac{n\hbar}{\mu}  p_\pm
                   -\kappa\frac{ r_\pm}{ r}\bigg)  |n,n-1,\pm(n-1)\ra.
\en
Using
$\la \vec r|n, l,m\ra=R_{nl} (r)Y_{lm}(\theta,\phi)$
the above equation takes the following form: 
\be
\bigg(\frac{ni\hbar}{\mu} (p_\pm)_{op}
                   +\kappa\frac{r_\pm}{ r}\bigg) R_{n,n-1}(r) Y_{n-1,\pm(n-1)}=0.
\en
With the help of $Y_{l,\pm l}\propto (r_\pm)^l/r^l$ and
\be
(p_\pm)_{op} (r_\pm)^l f(r)
=2\frac{\hbar}{i}\frac{\partial}{\partial r_\mp} (r_\pm)^l f(r)
=(r_\pm)^l \frac{\hbar}{i}\frac{r_\pm}{r}\frac{df(r)}{dr},
\en
we clearly see that the it is equivalent to Eq.~(\ref{eq: Rnn-1 equation}).

\subsubsection{Obtaining other $R_{nl}(r)$}

Once $R_{n,n-1}(r)$ is known, other $R_{n,l}(r)$ can be obtained readily by applying the second relation of the recursive relations given in Eq.~(\ref{eq: Rnlpm1 from Rnl}):
\be
\frac{R_{n,l-1}(r)}{r^{l-1}}
=\frac{n}{\sqrt{n^2-l^2}}
\bigg(\frac{(2l+1)l a_0}{Z}+\frac{l a_0 r}{Z}\frac{d}{dr}-r\bigg) \frac{R_{nl}(r)}{r^l}.
\en
For illustration, we apply the above equation on the
$l=n-1$ radial wave function and obtain 
\be
R_{n,n-2}(r)&=& c_n n(n-1)\sqrt{2n-1} \bigg(\frac{Zr}{a_0}\bigg)^{n-2} 
\bigg(1-\frac{Z}{n(n-1)a_0} r\bigg)e^{-\frac{Z}{na_0} r},
\en
and, sequentially, applying the relation on $l=n-2$, we have
\be
R_{n,n-3}(r)&=& \frac{1}{2} c_n\bigg(\frac{Z r}{a_0}\bigg)^{n-3}  (2n-3)(n-2)n^2\sqrt{(2n-1)(n-1)}
\non\\
&&\bigg(1-\frac{2Z r}{n(n-2) a_0}+\frac{2 Z^2 r^2}{n^2(6-7n+2n^2)a_0^2}\bigg) e^{-\frac{Zr}{n a_0}}.
\en
In principle, the procedure can be carried out to obtain all $R_{nl}(r)$.

It will be useful to show explicitly some of the radial wave functions obtained:
\be
R_{10}(r)&=&2 \bigg(\frac{Z}{a_0}\bigg)^{3/2} e^{-Zr/a_0}\cc
\non\\
R_{21}(r)&=&\frac{1}{\sqrt 3} \bigg(\frac{Z}{2 a_0}\bigg)^{3/2} \frac{Zr}{a_0}e^{-Zr/2a_0}\cc
\non\\
R_{20}(r)&=&2 \bigg(\frac{Z}{2 a_0}\bigg)^{3/2} \bigg(1-\frac{Zr}{2a_0}\bigg)e^{-Zr/2a_0}\cc
\non\\
R_{32}(r)&=&\frac{2\sqrt2}{27\sqrt 5} \bigg(\frac{Z}{3 a_0}\bigg)^{3/2} \bigg(\frac{Zr}{a_0}\bigg)^2 e^{-Zr/3a_0}\cc
\non\\
R_{31}(r)&=&\frac{4\sqrt2}{9} \bigg(\frac{Z}{3 a_0}\bigg)^{3/2} \bigg(\frac{Zr}{a_0}\bigg)
\bigg(1-\frac{Z r}{6 a_0}\bigg) e^{-Zr/3a_0}\cc
\non\\
R_{30}(r)&=&2 \bigg(\frac{Z}{3 a_0}\bigg)^{3/2} 
\bigg(1-\frac{2Z r}{3 a_0}+\frac{2(Z r)^2}{27 a^2_0}\bigg) e^{-Zr/3a_0},
\en
and compare them to those obtained by solving the Schr\"{o}dinger equation directly, see for example, \cite{Gasiorowicz}.
Indeed, it is clear that they are consistent with the results obtained in the two approaches.~\footnote{Note that there is a typo in the normalization factor of $R_{31}(r)$ in \cite{Gasiorowicz}.}
It is interesting that even the phase conventions match.

\subsection{The radial wave function $R_{nl}$ satisfies the radial Schr\"{o}dinger equation}

Applying the recursive relations, Eq.~(\ref{eq: Rnlpm1 from Rnl}), on $R_{n,l}/r^l$ can bring it to $R_{n,l\pm 1}/r^{l\pm 1}$ and suitably apply the relations again can bring them back to $R_{n,l}/r^l$. These procedures produce the following identities on $R_{nl}/r^l$,~\footnote{The additional factors
$1/(l+1)^2$ and $1/l^2$ are designed
to remove the coefficients of 
$d^2/dr^2$.}~\footnote{The equation is said to be factorizable~\cite{recursive1, recursive2, recursive3, recursive4}.}
\be
\frac{1}{(l+1)^2} \bigg((2l+3)(l+1)+(l+1) r\frac{d}{dr}-r\frac{Z}{a_0}\bigg)
\bigg(\frac{Z}{a_0r}+\frac{(l+1)}{r}\frac{d}{dr}\bigg) \frac{R_{nl}(r)}{r^l} 
\non\\
=-\frac{Z^2(n^2-(l+1)^2)}{n^2a^2_0 (l+1)^2} \frac{R_{nl}(r)}{r^l}\cc
\non\\
\frac{1}{l^2}\bigg(\frac{Z}{a_0 r}+\frac{l}{r}\frac{d}{dr}\bigg) 
\bigg((2l+1)l+l r\frac{d}{dr}-r\frac{Z}{a_0}\bigg)
\frac{R_{nl}(r)}{r^l} 
=-\frac{Z^2(n^2-l^2)}{n^2a^2_0 l^2}  \frac{R_{nl}(r)}{r^l},
\label{eq: Rnl=Rnl}
\en
giving,
\be
\bigg(\frac{d^2}{dr^2}+\frac{2(l+1)}{r}\frac{d}{dr}
+\frac{2 Z}{a_0 r}-\frac{Z^2}{(l+1)^2 a_0^2}\bigg)\frac{R_{nl}(r)}{r^l} 
&=&-\frac{Z^2(n^2-(l+1)^2)}{n^2a^2_0 (l+1)^2} \frac{R_{nl}(r)}{r^l},
\non\\
\bigg(\frac{d^2}{dr^2}+\frac{2(l+1)}{r}\frac{d}{dr}
+\frac{2 Z}{a_0 r}-\frac{Z^2}{l^2 a_0^2}\bigg)\frac{R_{nl}(r)}{r^l} 
&=&-\frac{Z^2(n^2-l^2)}{n^2a^2_0 l^2}  \frac{R_{nl}(r)}{r^l},
\en
or, equivalently,
\be
\bigg(\frac{d^2}{dr^2}+\frac{2(l+1)}{r}\frac{d}{dr}
+\frac{2 Z}{a_0 r}-\frac{Z^2}{n^2 a_0^2}\bigg)\frac{R_{nl}(r)}{r^l}=0.
\label{eq: Rn=Rn 1}
\en
It can be compared to the radial equation from the Schr\"{o}dinger equation, see for example, \cite{Gasiorowicz}
\be
\bigg(\frac{d^2}{dr^2}+\frac{2(l+1)}{r} \frac{d}{dr}+\frac{2Z}{a_0 r}+\frac{2\mu}{\hbar^2}E_n\bigg) \frac{R_{nl}}{r^l}=0.
\en
It is clear that Eq. (\ref{eq: Rn=Rn 1}) is same as
the above equation with
$E_n=-{Z^2\hbar^2}/{2\mu  a_0^2 n^2}$.

\subsection{Wave functions in the $E>0$ case}

We briefly discuss the $E>0$ case here.
For the $E>0$ case, as in the $E<0$ case, one can still define
\be
{\bf B}^{(\pm)}_i\equiv \frac{1}{2}({\bf L}_i\pm i {\bf A}')
\equiv \frac{1}{2}\bigg({\bf L}_i\pm i \sqrt{\frac{\mu}{2 E}}{\bf A}\bigg),
\en
which satisfy
\be
{[{\bf B}^{(+)}_i, {\bf B}^{(+)}_j]}_{\alpha\beta}&=&i\hbar \epsilon_{ijk} ({\bf B}^{(+)}_k)_{\alpha\beta},
\non\\
{[{\bf B}^{(-)}_i, {\bf B}^{(-)}_j]}_{\alpha\beta}&=&i\hbar \epsilon_{ijk} ({\bf B}^{(-)}_k)_{\alpha\beta},
\non\\
{[{\bf B}^{(+)}_i, {\bf B}^{(-)}_j]}_{\alpha\beta}&=&0,
\en
and
\be
( {\vec {\bf B}}^{(\pm)})^2
=\frac{1}{4}( {\vec {\bf L}}^2\pm i{\vec {\bf L}}\cdot{\vec {\bf A}'}\pm i{\vec {\bf A}'}\cdot{\vec {\bf L}}-{\vec {\bf A}}^{\prime 2})
=\frac{1}{4}( {\vec {\bf L}}^2-{\vec {\bf A}}^{\prime 2}) =\frac{1}{4}(-\hbar^2-\frac{\mu}{2 E}\kappa^2){\bf 1}.
 \label{eq: B(pm)2 E>0}
\en
However, now the situation is different. The above relations cannot be used to obtain energy eigenvalue as in the $E<0$ case. 
The reason is that $ {\vec {\bf B}}^{(\pm)}$ are no longer hermitian. 
The usual procedure of obtaining quantum numbers in ${\vec {\bf L}}^2$ and ${\bf L}_z$ 
cannot be used in $ ({\vec {\bf B}}^{(\pm)})^2$ and $ {\bf B}^{(\pm)}_z$. 
For example, the above equation show that $( {\vec {\bf B}}^{(\pm)})^2$ are negative matrices
and the usual steps in quantizing angular momentum break down in quantizing this system.

Nevertheless some results obtained in the previous section can still be used.
In particular the recursive relations similar to Eq. (\ref{eq: Rnlpm1 from Rnl}) can be used to find the wave function in the $E>0$ case.
Replacing $n$ by $i\nu$ in Eq. (\ref{eq: Rnlpm1 from Rnl}), 
we have the following recursive relations
\be
-\frac{Z\sqrt{\nu^2+(l+1)^2}}{\nu a_0}
\frac{R_{l+1}(r)}{r^{l+1}}
&=&\bigg(\frac{(l+1)}{r}\frac{d}{dr}+\frac{Z}{a_0 r}\bigg) \frac{R_{l}(r)}{r^l},
\non\\
\frac{Z\sqrt{\nu^2+l^2}}{\nu a_0}\frac{R_{l-1}(r)}{r^{l-1}}
&=&\bigg((2l+1)l+l r\frac{d}{dr}-r\frac{Z}{a_0}\bigg) \frac{R_{l}(r)}{r^l},
\label{eq: Rnlpm1 from Rnl E>0}
\en
which lead to
\be
\bigg(\frac{d^2}{dr^2}+\frac{2(l+1)}{r}\frac{d}{dr}
+\frac{2 Z}{a_0 r}-\frac{Z^2}{(l+1)^2 a_0^2}\bigg)\frac{R_{l}(r)}{r^l} 
&=&\frac{Z^2(-\nu^2-(l+1)^2)}{\nu^2a^2_0 (l+1)^2} \frac{R_{l}(r)}{r^l},
\non\\
\bigg(\frac{d^2}{dr^2}+\frac{2(l+1)}{r}\frac{d}{dr}
+\frac{2 Z}{a_0 r}-\frac{Z^2}{l^2 a_0^2}\bigg)\frac{R_{l}(r)}{r^l}
&=&\frac{Z^2(-\nu^2-l^2)}{\nu^2a^2_0 l^2}  \frac{R_{l}(r)}{r^l},
\label{eq: Rnl=Rnl E>0}
\en
or, simply,
\be
\bigg(\frac{d^2}{dr^2}+\frac{2(l+1)}{r}\frac{d}{dr}
+\frac{2 Z}{a_0 r}+\frac{Z^2}{\nu^2 a_0^2}\bigg)\frac{R_{l}(r)}{r^l}=0.
\en
The above equation can match to the Schr\"{o}dinger equation (with $E>0$):
\be
\bigg(\frac{d^2}{dr^2}+\frac{2(l+1)}{r} \frac{d}{dr}+\frac{2Z}{a_0 r}+\frac{2\mu}{\hbar^2}E\bigg) \frac{R_{l}}{r^l}=0,
\en
by taking 
\be
\nu=\frac{Z\hbar}{\sqrt{2\mu E}  a_0}=\frac{Z}{a_0 k},
\en
with $k\equiv\sqrt{2\mu E}/\hbar$.
Solving the Schr\"{o}dinger equation now reduce to finding functions that satisfy the recursive relations in Eq.~(\ref{eq: Rnl=Rnl E>0}).

From \cite{AS}, we find that the Coulomb functions, $u_l(\eta,\rho)=F_l(\eta,\rho)$ and $G_l(\eta,\rho)$, have the following recursive relations:
\be
l\frac{d}{d\rho} u_l(\eta,\rho)+\bigg(\frac{l^2}{\rho}+\eta\bigg)u_l&=&\sqrt{l^2+\eta^2} u_{l-1},
\non\\
(l+1)\frac{d}{d\rho} u_l(\eta,\rho)-\bigg(\frac{(l+1)^2}{\rho}+\eta\bigg)u_l&=&-\sqrt{(l+1)^2+\eta^2} u_{l+1},
\en
or, equivalently,
\be
l\rho\frac{d}{d\rho} \frac{u_l}{\rho^{l+1}}+l(2l+1) \frac{u_l}{\rho^{l+1}}+\eta\rho\frac{u_l}{\rho^{l+1}}&=&\sqrt{l^2+\eta^2} \frac{u_{l-1}}{\rho^{l}},
\non\\
\frac{(l+1)}{\rho}\frac{d}{d\rho} \frac{u_l}{\rho^{l+1}}-\frac{\eta}{\rho} \frac{u_l}{\rho^{l+1}}&=&-\sqrt{(l+1)^2+\eta^2} \frac{u_{l+1}}{\rho^{l+2}}.
\en
Compare the above relations with those in Eq. (\ref{eq: Rnlpm1 from Rnl E>0}), we see that by taking
\be
\eta=-\nu=-\frac{Z}{a_0 k},
\quad
\rho= k r,
\en
the following $R_l/r^l$, 
\be
\frac{R_l(r)}{r^l}= a \frac{F_l(-Z/a_0 k,kr)}{r^{l+1}}+b \frac{G_l(-Z/a_0 k,kr)}{r^{l+1}},
\en
with constants $a$, $b$,
satisfies the recursive relations in Eq. (\ref{eq: Rnlpm1 from Rnl E>0}).
Indeed, the above result on $R_l$ is consistent with those obtained by solving the Schr\"{o}dinger equation directly, see for example, \cite{Coulomb}.

\section{Conclusions}

In this work we follow the 
Pauli method of quantizing the Hydrogen system using the Runge-Lenz vector. 
Since the Runge-Lenz vector is a vector and it commutes with the Hamiltonian, 
it is natural to use it to connect energy eigenstate $|n,l,m\rangle$ with other degenerate states $|n, l\pm 1,m'\rangle$.
In this work, we found the recursive relations for the radial wave functions. 
They are consistent with the results found in \cite{recursive1, recursive2, recursive3, recursive4} using the factorization method.
Nevertheless we believe that as the present approach makes good use of the properties of the Runge-Lenz vector, 
which is a conserved vector,
it is the most natural way to obtain the recursive relations.
The wave functions of the whole spectrum can be obtained easily.
These radial wave functions are shown to satisfy the Schr\"{o}dinger equation.
In addition, using the recursive relations the wave functions in the $E>0$ case can also be verified.
As in the Pauli analysis, group theory plays a prominent role in this analysis, while the rest of the derivations are mostly elementary.

\section*{Acknowledgments}

The authors are grateful to Chung-Wen Kao for discussions.
This research was supported in part by the Ministry of Science and Technology of R.O.C. under Grant Nos. 103-2112-M-033-002-MY3 and 106-2112-M-033-004-MY3.

\appendix

\section{Obtaining the matrix elements of $B^{(\pm)}_-$ as shown in Eq. (\ref{eq: group})}

In Sec.~\ref{connecting}, we need to evaluate the following matrix elements of $B^{(\pm)}_-$:
\be
\la n, l',l-1| B^{(+)}_-|n, l,l\ra
&=&\la l', l-1|b, m^{(+)}-1, b, m^{(-)} \ra \sqrt{(b+m^{(+)})(b-m^{(+)}+1)} \hbar
\non\\
&&
\la b, m^{(+)}, b, m^{(-)} |l,l\ra,
\non\\
\la n, l',l-1| B^{(-)}_-|n, l,l\ra
&=&\la l', l-1|b, m^{(+)}, b, m^{(-)}-1 \ra \sqrt{(b+m^{(-)})(b-m^{(-)}+1)} \hbar
\non\\
&&
\la b, m^{(+)}, b, m^{(-)} |l,l\ra,
\label{eq: A1}
\en
with $m^{(\pm)}$ summed.
First we note that from the known symmetrical properties of Clebsch-Gordan coefficients,
\be
\la l', l-1|b, m^{(+)}, b, m^{(-)}-1 \ra &=&(-)^{2b-l'}\la l', l-1|b, m^{(-)}-1, b, m^{(+)} \ra,
\non\\
\la b, m^{(+)}, b, m^{(-)} |l,l\ra&=&(-)^{2b-l}\la b, m^{(-)}, b, m^{(+)} |l,l\ra,
\en
we have
\be
\la n, l',l-1| B^{(-)}_-|n, l,l\ra=(-)^{l-l'}\la n, l',l-1| B^{(+)}_-|n, l,l\ra,
\label{eq: B- B+}
\en
for integers $l,l'$, which is certainly the case here.

Using the known formula in the study of addition of angular momenta~\cite{angularmomentum} (see, also~\cite{rose}),
\be
\la j'_1j'_2;j'm'|T^\mu_k(1)|j_1j_2;jm\ra
&=&
(-)^{-j'_1-j_2-j-k}\la j,m, k,\mu|j',m'\ra (2j+1)^{1/2}(2j'_1+1)^{1/2} 
\non\\
&&
\bigg\{
\begin{array}{ccc}
j_2 &j_1 &j\\
k    &j'    &j'_1
\end{array}
\bigg\}
\la j'_1|| T_k(1) ||j_1\ra,
\en
where $\{\,\,\}$ is the Wigner 6-$j$ symbol and $\la j'_1|| T_k(1) ||j_1\ra$ is the reduced matrix element,
we must have
\be
\la n,l',l-1|B^{(+)}_-|n,l,l\ra
&=&
(-)^{-2b-l+1}\la l,l, 1,-1|l'\,l-1\ra (2l+1)^{1/2}(2b+1)^{1/2} 
\bigg\{
\begin{array}{ccc}
b &b &l\\
1    &l'    &b
\end{array}
\bigg\}
\non\\
&&\times
\la b|| B^{(+)} ||b\ra,
\en
with $j_1=j_2=j'_1=j'_2=b=(n-1)/2$. For $l'=l,l\pm 1$, the above formula and Eq. (\ref{eq: B- B+}) give
\be
\la n,l-1,l-1|B^{(+)}_-|n,l,l\ra
&=&-\sqrt{\frac{l(n^2-l^2)}{(2l+1)(n^2-1)}}
\la b|| B^{(+)} ||b\ra
=-\la n,l-1,l-1|B^{(-)}_-|n,l,l\ra,
\non\\
\la n,l+1,l-1|B^{(+)}_-|n,l,l\ra
&=&\sqrt{\frac{n^2-(l+1)^2}{(2l+1)(2l+3)(n^2-1)}} 
\la b|| B^{(+)} ||b\ra
=-\la n, l+1,l-1|B^{(-)}_-| n,l,l\ra,
\non\\
\la n,l,l-1|B^{(+)}_-|n, l,l\ra
&=&\sqrt{\frac{l}{n^2-1}}
\la b|| B^{(+)} ||b\ra
=\la n, l,l-1|B^{(-)}_-| n,l,l\ra.
\label{eq: A6}
\en

The reduced matrix element $\la b|| B^{(+)}|| b\ra$ can be determined by using
\be
\la n,l,l-1|B^{(+)}_-|n, l,l\ra+\la n,l,l-1|B^{(-)}_-|n, l,l\ra
=\la n,l,l-1|L_-|n, l,l\ra=\sqrt{2l}\,\hbar,
\en
and the third relation in Eq. (\ref{eq: A6}), giving
\be
\la b|| B^{(+)} ||b\ra=\sqrt{\frac{n^2-1}{2}}\hbar,
\en
and, consequently, we obtain
\be
\la n, l-1,l-1| B^{(\pm)}_-|n, l,l\ra 
&=&\mp \sqrt{\frac{l(n^2-l^2)}{2(2l+1)}}\hbar,
\non\\
\la n, l+1,l-1| B^{(\pm)}_-|n, l,l\ra 
&=&\pm \sqrt{\frac{n^2-(l+1)^2}{2(2l+1)(2l+3)}}\hbar,
\non\\
\la n, l,l-1| B^{(\pm)}_-|n, l,l\ra 
&=&\sqrt{\frac{l}{2}}\hbar.
\en
These are the results shown in Eq. (\ref{eq: group}). They follow from group theory.

Note that the above results also hold for the $l=0$ case, where the equation implies that $\la n, l-1,l-1| B^{(\pm)}_-|n, l,l\ra$ and $\la n, l,l-1| B^{(\pm)}_-|n, l,l\ra$ are vanishing as they should.
Furthermore, in the above derivation we encounter $1/(n^2-1)^{1/2}$ factors in Eq. (\ref{eq: A6}), and, hence, $n$ needs to be greater than 1.
Nevertheless, using Eq. (\ref{eq: A1}) and by direct computation, it is easy to check that the above equation also holds for the $n=1$ case.

\end{document}